\documentclass{elsart}
\journal{Physics Letters B}

\usepackage{graphicx}

\usepackage{amssymb}

\begin{document}

\begin{frontmatter}

\hfill{UK/06-11}



\title{On the Locality and Scaling of Overlap Fermions at Coarse Lattice Spacings}


\author[UK]{Terrence Draper,}
\author[JLAB]{Nilmani Mathur,}
\author[Zhe]{Jianbo Zhang,}
\author[UK]{Andrei Alexandru,}
\author[IHEP]{Ying Chen,}
\author[UK]{Shao-Jing Dong,}
\author[UK]{Ivan Horv\'{a}th,}
\author[GW]{Frank X. Lee,}
\author[UK]{Keh-Fei Liu,}
\author[Ham]{Sonali Tamhankar}
\address[UK]{Department of Physics and Astronomy, 
             University of Kentucky, 
             Lexington, KY 40506, USA}
\address[JLAB]{Jefferson Lab, 
              12000 Jefferson Avenue, 
              Newport News, VA 23606, USA}
\address[Zhe]{Department of Physics,
              Zhejiang University, 
              Hangzhou, Zhejiang 310027, China}
\address[IHEP]{Institute of High Energy Physics, 
               Chinese Academy of Sciences, 
	       Beijing 100049, China}
\address[GW]{Center for Nuclear Studies, 
             Department of Physics, 
             George Washington Univ.,
             Washington, DC 20052, USA}
\address[Ham]{Department of Physics, 
              Hamline University,
              St. Paul, MN 55104, USA}



\begin{abstract}
The overlap fermion offers the considerable advantage of exact chiral symmetry
on the lattice, but is numerically intensive.  This can be made affordable
while still providing large lattice volumes, by using coarse lattice spacing,
given that good scaling and localization properties are established. Here,
using overlap fermions on quenched Iwasaki gauge configurations, we demonstrate
directly that, with appropriate choice of negative Wilson's mass, the overlap
Dirac operator's range is comfortably small in lattice units for each of the
lattice spacings 0.20 fm, 0.17 fm, and 0.13 fm (and scales to zero in physical
units in the continuum limit).  In particular, our direct results contradict
recent speculation that an inverse lattice spacing of $1\,{\rm GeV}$ is too low
to have satisfactory localization.  Furthermore, hadronic masses (available on
the two coarser lattices) scale very well.
\end{abstract}

\end{frontmatter}


\section{Locality}

In the last several years the use of the overlap lattice fermion has become
more popular because the conceptual and technical clarity that results from its
exact chiral symmetry is seen to overcome its superficially higher
computational cost as compared to conventional lattice formulations of the
fermion.  Furthermore, the disparity in numerical intensity can be mitigated by
using coarse lattice spacing, given that good scaling and localization
properties are established.

Hern\'{a}ndez, Jansen, and L\"{u}scher~\cite{Hernandez:1998et} have shown
numerically that Neuberger's overlap operator is local on pure-glue backgrounds
(using the Wilson gauge action on fine lattices).  The decay rate (inverse of
the range) $r_{\rm ov}^{-1}$ of the overlap kernel can be interpreted as the
mass of unphysical degrees of freedom.  Recently, Golterman, Shamir, and
Svetitsky~\cite{Golterman:2005fe} have argued that, in practice, these can be
dangerously light.  They provide an indirect estimate for pure-gauge ensembles
(for Wilson, Iwasaki, and DBW2 gauge actions) at different lattice spacings.
For cutoff $a^{-1}\sim 2\,{\rm GeV}$ they estimate $r_{\rm ov}^{-1}\approx 0.4
\times 2 {\rm GeV} = 800\,{\rm MeV}$, which is acceptably large.  For
$a^{-1}\sim 1\,{\rm GeV}$, however, they raise an alarm with their estimate of
$r_{\rm ov}^{-1}\approx 0.25 \times 1 {\rm GeV} = 250\,{\rm MeV}$ for the
Wilson case, $270\,{\rm MeV}$ for Iwasaki, and $320\,{\rm MeV}$ for DBW2.

However, their estimate is indirect, and follows a chain of approximations at
the end of which they assume that $r_{\rm ov}^{-1}\approx \lambda_{c}^{-1}$,
where $\lambda_{c}$ is the ``mobility edge'' of a Dirac spectrum.  Here we come
to a very different conclusion using the most direct approach which takes
advantage of the fact that it is not expensive to {\it calculate}, rather than
{\it estimate}, $<D_{\rm ov}(x,y)>$.  (By comparison, it would be much more
expensive to calculate the overlap propagator $D_{\rm ov}^{-1}$ which must be
computed by nested iterations.)

In contrast to the results of Golterman, Shamir, and
Svetitsky~\cite{Golterman:2005fe}, who speculate that overlap simulations with
a cutoff of $1\,{\rm GeV}$ (such as~\cite{Chen:2003im}) might have a range as
long as 4 lattice units, here we show directly that such is not the case; with
appropriate choice of negative Wilson's mass, the range is about 1 lattice unit
(in Euclidean distance or 2 units of ``taxi-driver'' distance).  All is well.

A preliminary version of these results was presented in~\cite{Draper:2005mh}.
Durr, Hoelbling, and Wenger~\cite{Durr:2005an} came to similar conclusions
using the overlap operator with a UV filtered (``thick link'') Wilson kernel.

\subsection{Lattice Details}

We use the renormalization-group-improved Iwasaki~\cite{Iwasaki:1985we} gauge
action, on three different lattices; for each, the lattice size, lattice
spacing, and number of configurations used are tabulated in
Table~\ref{Table:lattices}.

\begin{table}[htb] 
  \begin{center}
    \begin{tabular}{lllr}
      $N_s \times N_t$   & $a (r_0)$          & $a (f_{\pi})$     &  $N_{\rm cfg}$ \\
      \hline				                   
      $16^3\times 28$    & $0.175\,{\rm fm}$  & $0.20\,{\rm fm}$  & $300/10$ \\
      $20^3\times 32$    & $0.153\,{\rm fm}$  & $0.17\,{\rm fm}$  & $ 98/10$ \\
      $28^3\times 44$    & $0.113\,{\rm fm}$  & $0.13\,{\rm fm}$  & $  0/10$ \\
      \hline
    \end{tabular}
    \caption{\label{Table:lattices} Lattice size, lattice spacing (as set by
             the Sommer parameter $r_0$~\cite{Sommer:1993ce}), lattice spacing
             (as set by $f_{\pi}(m_{\pi})$~\cite{Chen:2003im}), and number of
             configurations (for scaling/locality). }
  \end{center}
\end{table}

For the associated scaling study of hadron masses, we use the overlap
fermion~\cite{Neuberger:1998fp,Narayanan:1995gw} and massive overlap
operator~\cite{Alexandrou:2000kj,Capitani:2000wi,Hernandez:2001yn,Dong:2001fm}
\begin{equation}
  D(m_0) = (\rho + \frac{m_0a}{2}) + (\rho - \frac{m_0a}{2} ) \gamma_5 \epsilon (H)
\end{equation}
where $\epsilon (H) = H /\sqrt{H^2}$, $H = \gamma_5 D_w$, and $D_w$ is the
usual Wilson fermion operator, except with a negative mass parameter $-\rho =
1/2\kappa -4$ in which $\kappa_c < \kappa < 0.25$; we take $\kappa = 0.19$ in
our calculation which corresponds to $\rho = 1.368$.  For the locality study,
we set $m_0=0$ to look at the properties of the massless operator $D(0)$.
Complete details are described in~\cite{Chen:2003im}.

\subsection{Locality as Measured by Taxi-Driver Distance}

It has been convenient to discuss locality in terms of a ``taxi-driver''
distance~\cite{Hernandez:1998et}, rather than the standard Euclidean distance.
\begin{equation}
  r_{\rm TD} \equiv || x-y||_1 = \sum_{\mu} |x_\mu - y_\mu|
\end{equation}
The locality of the overlap operator is then studied by plotting the quantity
$\mathcal{D}_{\rm max}(r)$ ($f(r)$ in the notation of~\cite{Hernandez:1998et}) as
a function of the taxi-driver distance for a localized source,
$\psi_{\alpha}(x)=\delta(x)\delta_{\alpha\beta}$ for fixed Dirac-color index
$\beta$.
\begin{equation} \label{Taxi_def}
  \mathcal{D}_{\rm max}(r) \equiv \max \{ ||D\psi(x)|| \,\, | \,\, \sum_{\mu} x_{\mu}=r \}
\end{equation}
For large $r$, the Dirac-overlap operator decays exponentially with decay rate
$\nu=r_{\rm ov}^{-1}$, where $r_{\rm ov}$ is the range (characteristic decay
distance) measured in lattice units.

\subsection{Results}

\begin{figure}[htb]
  \vspace{0cm}
  \begin{center}
  \includegraphics[angle=0,width=0.8\hsize]{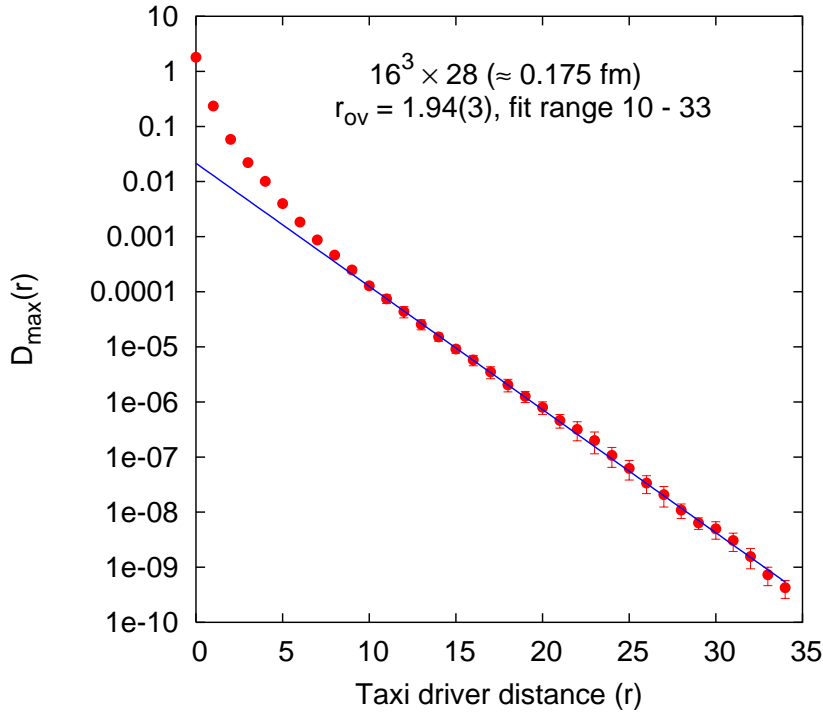}\\%
  \vspace{0cm}
  \caption{\label{Fig:Taxi_16} For the $16^3\times 28$ lattice, the expectation
           value of $\mathcal{D}_{\rm max}(r)$ as a function of the taxi-driver
           distance, $r$.  For large $r$, $\mathcal{D}_{\rm max}(r)$ falls
           exponentially, with range $r_{\rm ov}$.  The fitted value of $r_{\rm
           ov}$ is shown with the chosen fit interval.}
  \vspace{0cm}
  \end{center}
\end{figure}

In Fig.~\ref{Fig:Taxi_16}, we plot $\mathcal{D}_{\rm max}(r)$ as a function of
taxi-driver distance for the $16^3\times 28$ lattice.  At large distances, we
fit to an exponentially decreasing function to extract the range $r_{\rm ov}$.

\begin{figure}[htb]
  \vspace{0cm}
  \begin{center}
  \includegraphics[angle=0,width=0.8\hsize]{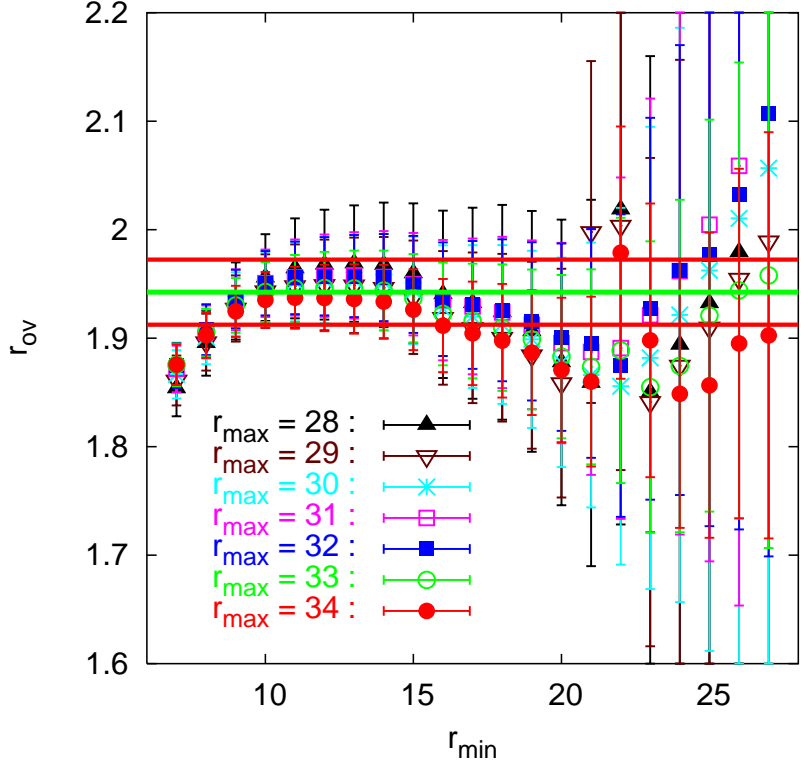}%
  \vspace{0cm}
  \caption{\label{Fig:Taxi_16_choice} For the $16^3\times 28$ lattice, the
           fitted values of $r_{\rm ov}$ for various choices of fit interval
           ($r_{\rm min}$,$r_{\rm max}$) are displayed as a function of $r_{\rm
           min}$ for each $r_{\rm max}$, to help choose the best fit.}
  \vspace{0.5cm}
  \end{center}
\end{figure}

To help choose the optimal fit interval, we display in
Fig.~\ref{Fig:Taxi_16_choice} the fitted values of $r_{\rm ov}$ for various
choices of fit interval ($r_{\rm min}$,$r_{\rm max}$) as a function of $r_{\rm
min}$ for each $r_{\rm max}$.  In analogy with one's experience fitting the
ground state mass of a two-point hadronic correlation function, for
sufficiently large $r_{\rm max}$ one would expect to see the fitted value of
the range $r_{\rm ov}$ {\it increase\/} (since $1/r_{\rm ov}$ is a mass) toward
a plateau as $r_{\rm min}$ is increased.  And as $r_{\rm min}$ is increased
further one would expect to see the plateau disappear into noise.  This is what
we see in Fig.~\ref{Fig:Taxi_16_choice}.  We choose the least value of $r_{\rm
min}$ for which the plateau is established, and a value $r_{\rm max}$ for which
the fit errors fairly represent the dispersion of results from various other
acceptable fit intervals.

\begin{figure}[htb]
  \vspace{0cm}
  \begin{center}
  \includegraphics[angle=0,width=0.52\hsize]{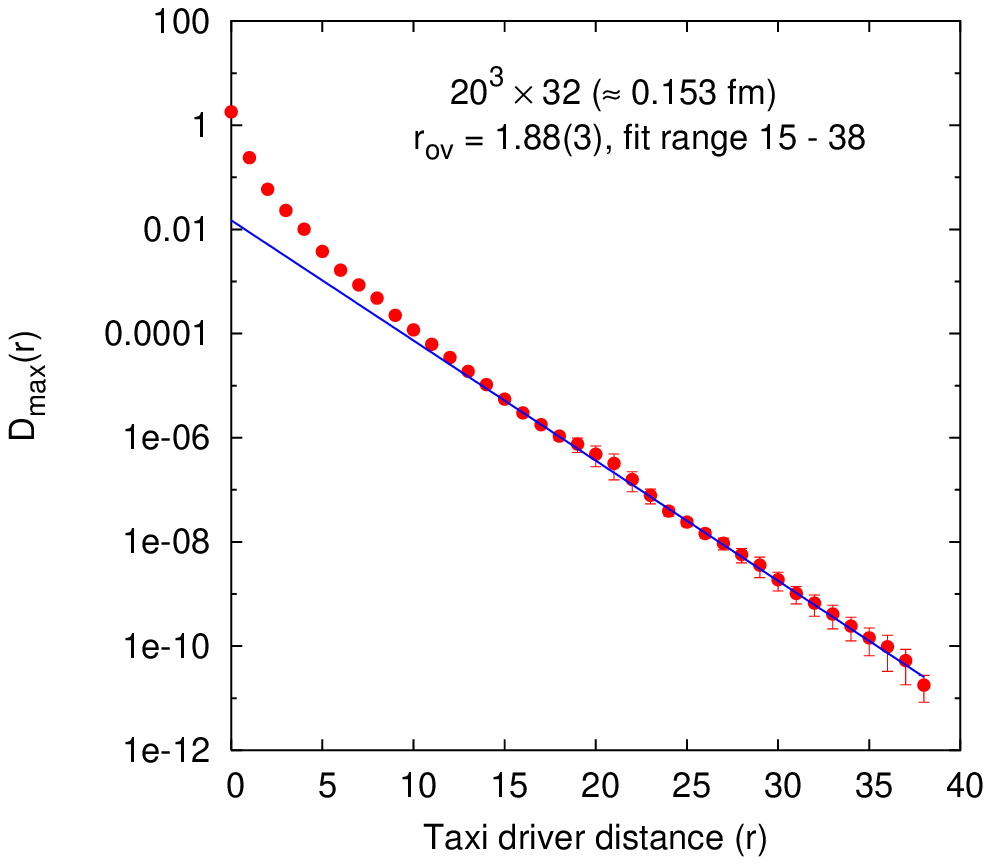}%
  \includegraphics[angle=0,width=0.48\hsize]{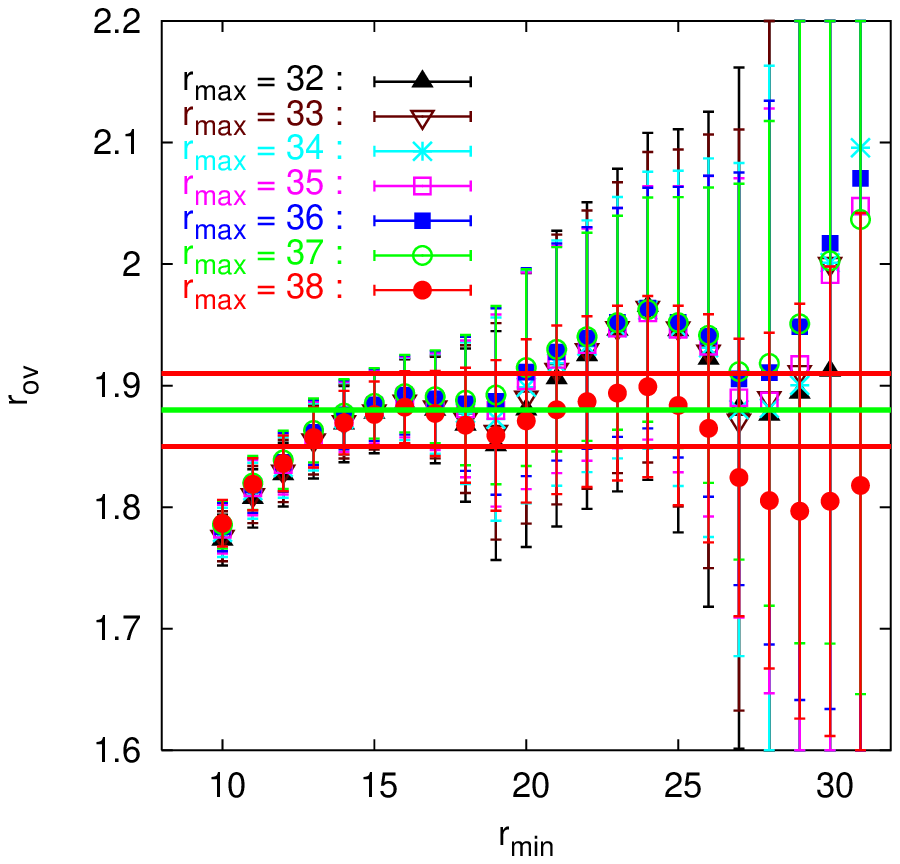}%
  \vspace{0cm}
  \caption{\label{Fig:Taxi_20} Same as figures~\ref{Fig:Taxi_16}
           and~\ref{Fig:Taxi_16_choice} but for the $20^3\times 32$ lattice.}
  \vspace{0.5cm}
  \end{center}
\end{figure}

In Fig.~\ref{Fig:Taxi_20} (left pane), we plot $\mathcal{D}_{\rm max}(r)$ as a
function of taxi-driver distance for the $20^3\times 32$ lattice, showing the
fit of $r_{\rm ov}$ for this lattice.  The fit interval is chosen upon analysis
of Fig.~\ref{Fig:Taxi_20} (right pane).

\begin{figure}[htb]
  \vspace{0cm}
  \begin{center}
  \includegraphics[angle=0,width=0.52\hsize]{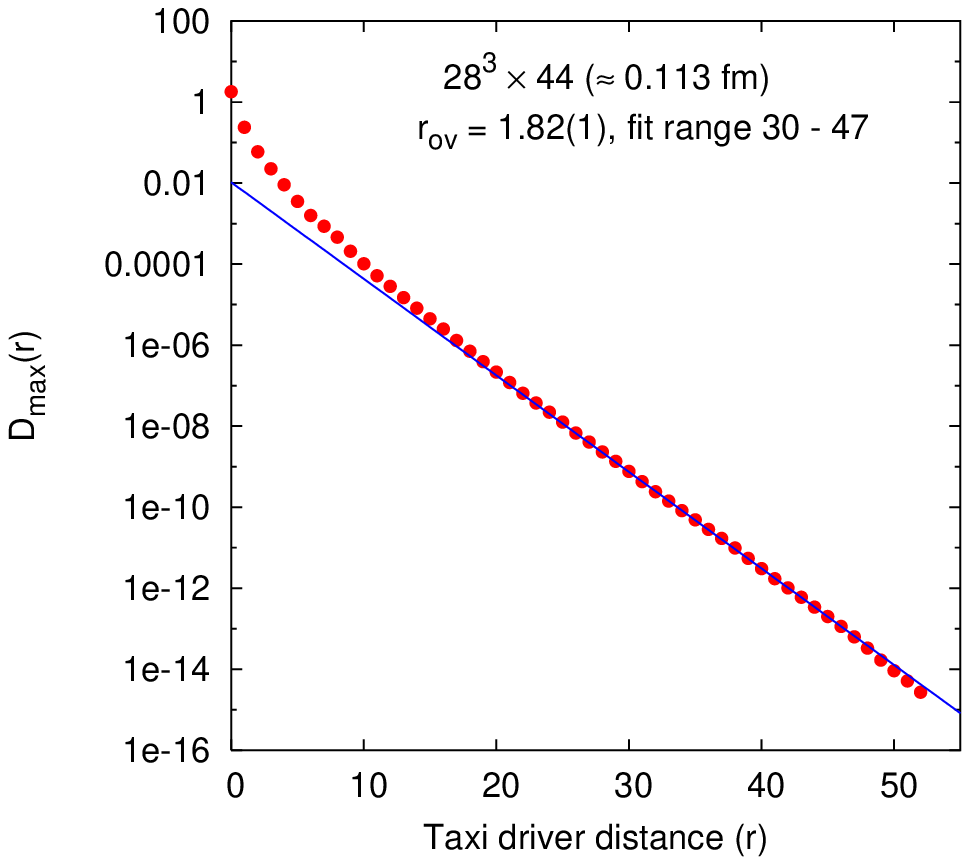}%
  \includegraphics[angle=0,width=0.48\hsize]{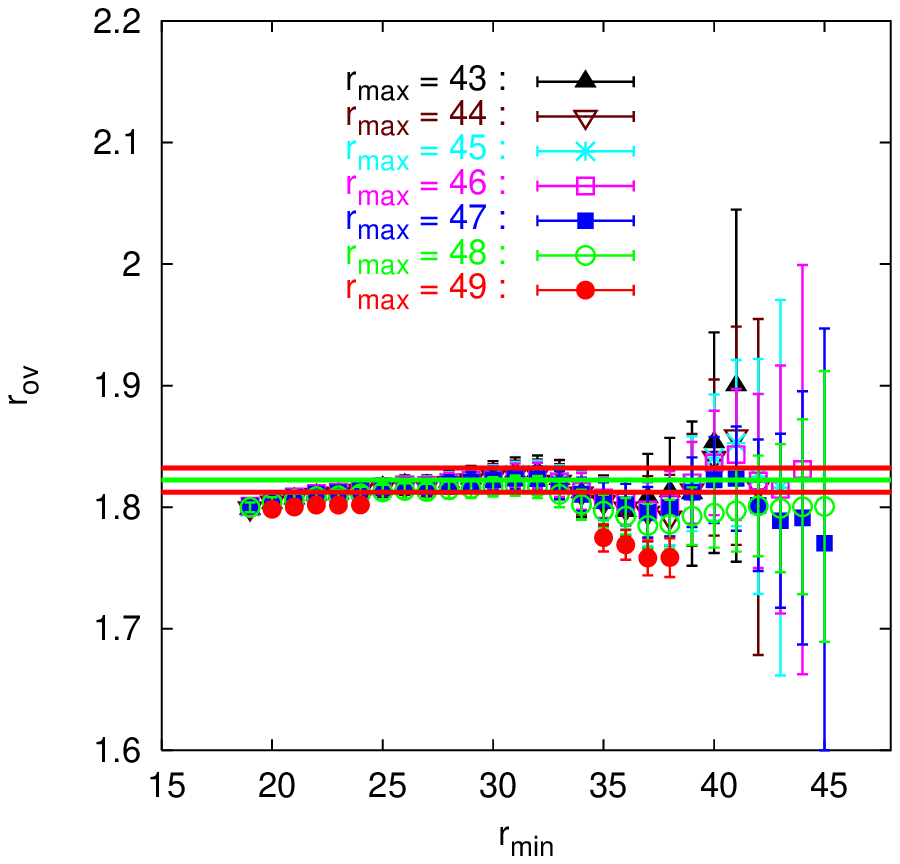}%
  \vspace{0cm}
  \caption{\label{Fig:Taxi_28} Same as figures~\ref{Fig:Taxi_16}
           and~\ref{Fig:Taxi_16_choice} but for the $28^3\times 44$ lattice.}
  \vspace{0cm}
  \end{center}
\end{figure}

Finally, in Fig.~\ref{Fig:Taxi_28} (left pane), we plot $\mathcal{D}_{\rm
max}(r)$ as a function of taxi-driver distance for the $28^3\times 44$ lattice,
showing the fit of $r_{\rm ov}$ for this lattice.  Again, the fit interval is
chosen from analysis of an ``$r_{\rm min}$'' plot, Fig.~\ref{Fig:Taxi_28}
(right pane).

\pagebreak[1]

Our fitted values of $r_{\rm ov}$ for each of our three lattices are tabulated
in Table~\ref{Table:locality}.  We note that our results are similar to the
results of Hern\'{a}ndez, Jansen and L\"{u}scher~\cite{Hernandez:1998et}
obtained on finer lattices for the overlap operator ($\rho=1.4$) with Wilson
action at $\beta=6.0$, $6.2$, and $6.4$; they find $r_{\rm
ov}=\nu^{-1}=1/0.49\approx 2.0$.

\begin{figure}[ht]
  \vspace{0cm}
  \begin{center}
  \includegraphics[angle=0,width=0.8\hsize]{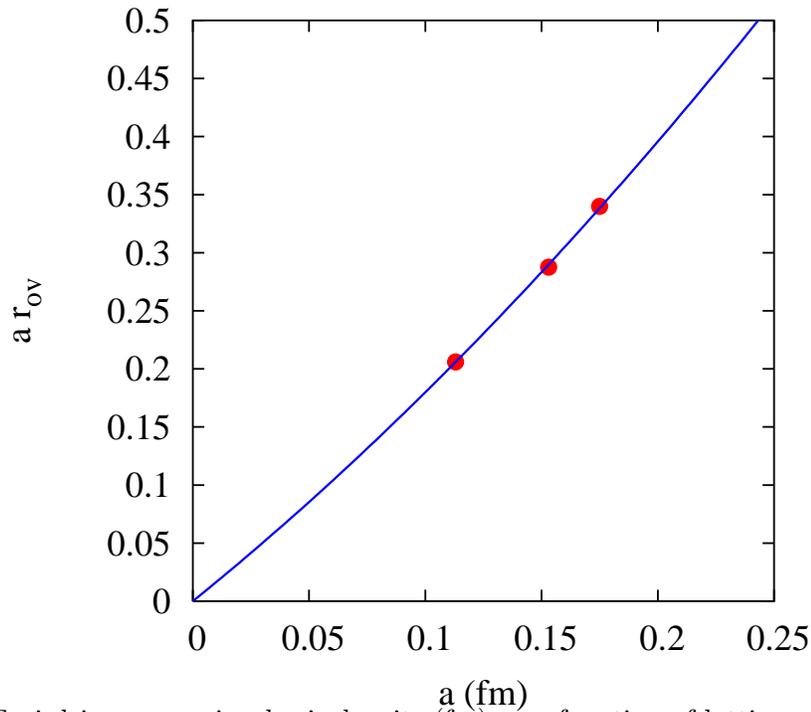}
  \vspace{-0.5cm}
  \caption{\label{Fig:spacing} Taxi driver range in physical units (fm) as a
           function of lattice spacing ($r_0$ scale).  The range is small even
           at coarse lattice spacing and trends to zero in the continuum
           limit.}
  \vspace{0cm}
  \end{center}
\end{figure}

\begin{table}[ht]
  \begin{center}
    \begin{tabular}{ll}
      $a(r_0)$          &  $r_{\rm ov}$ \\
      \hline
      $0.175\,{\rm fm}$ & $1.94(3)$ \\
      $0.153\,{\rm fm}$ & $1.88(3)$ \\
      $0.113\,{\rm fm}$ & $1.82(1)$ \\
      \hline
    \end{tabular}
    \caption{\label{Table:locality} The range (taxi driver metric) for three
             lattice spacings.  It is less than two lattice units on a lattice
             as coarse as $0.20\,{\rm fm}$ ($f_{\pi}$ scale).}
  \end{center}
\end{table}

Furthermore, it is gratifying to see that even for our coarsest lattice,
$0.20\,{\rm fm}$ ($f_{\pi}$ scale), the {\em measured\/} range is less than two
lattice units.  In Fig.~\ref{Fig:spacing} we plot the range in physical units
as a function of lattice spacing ($r_0$ scale).  It trends to zero in the
continuum limit; a quadratic fit restrained to go through the origin has a
satisfactory $\chi^2/dof=0.31$.

\subsubsection{Wrap-around Effects}

In this section, we delve further into some of the details used in extracting
the overlap range.  In particular, the raw data needs to be cut in order to
reduce wrap-around effects to an acceptable and conservative level.

\begin{figure}[ht]
  \vspace{0cm}
  \begin{center}
  \includegraphics[angle=0,width=0.7\hsize]{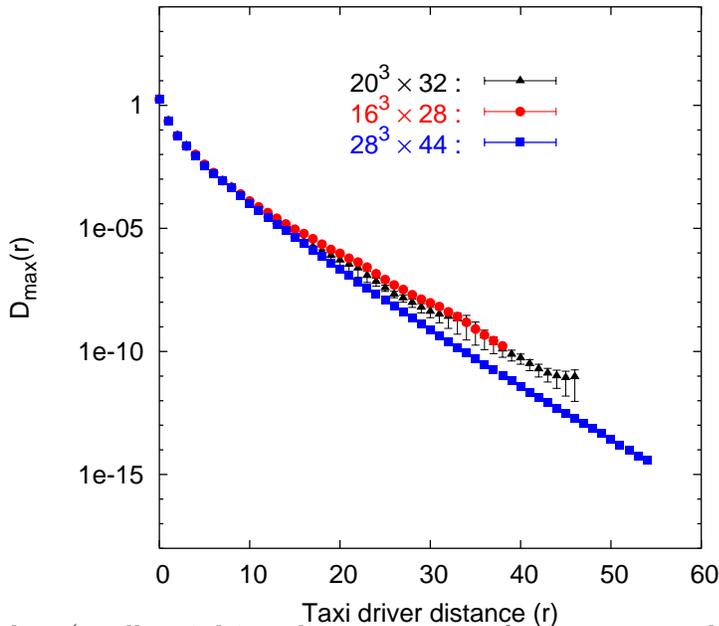}
  \vspace{-0.5cm}
  \caption{\label{Fig:raw} Raw data (at all taxi-driver distances except for
           some at very large distances on the largest lattice where the
           precision is inadequate).  This includes distances contaminated by
           wrap-around effects.}

  \vspace{0cm}
  \end{center}
\end{figure}

Our raw data is displayed in Fig.~\ref{Fig:raw}.  This includes data at all
taxi-driver distances except for some at very large distances on the largest
lattice where the numerical precision used in calculating $D$ may be
inadequate.

This raw data includes values at large distances which are contaminated by
wrap-around effects, due to the finite volume of the lattices; these must be
accounted for in order to extract the range of the overlap operator.  To
quantify the extent of the wrap-around effects, we consider first a subset of
data restricted to be along the $t$ axis.  If the lattice were infinite in time
extent, one could extract the inverse range $\nu\equiv r_{\rm ov}^{-1}$ by
defining an ``effective inverse range'' $\nu_{\rm exp}(t)\equiv \ln
\left(D\psi(t-1)/D\psi(t)\right)$ (like an 
``effective mass''~\cite{Bernard:1982hh}), based on the fit-model
$D\psi(t)\sim\exp{(-\nu t)}$ for the exponential tail.  At large time $t$, a
``plateau'' is seen where $\nu_{\rm exp}(t)$ is independent of $t$.  However,
for a finite lattice of time extent $T$, the exponential fit model breaks down
due to additive contributions $\sim
\exp{\left(-\nu (T-t)\right)}$ from wrap around and, instead, the fit model
$D\psi(t)\sim \cosh{(-\nu (t-T/2))}$ is appropriate for times $t$ near the
middle of the lattice ($ 1 \ll t \ll T$).  But still, one can define another
``effective inverse range'' $\nu_{\rm cosh}(t)$, appropriate for the new fit
model; this forms a plateau in the middle of the periodic lattice.  For small
$t$ on a periodic finite lattice, $\nu_{\rm exp}(t)
\approx \nu_{\rm cosh}(t)$ with exponentially small difference as the
wrap-around effects are negligible.  But near the middle of the lattice, while
the cosh effective range forms a plateau, the exponential effective range
deviates exponentially due to the wrap-around effects.

\begin{figure}[ht]
  \vspace{0cm} \begin{center}
  \includegraphics[angle=0,width=0.5\hsize]{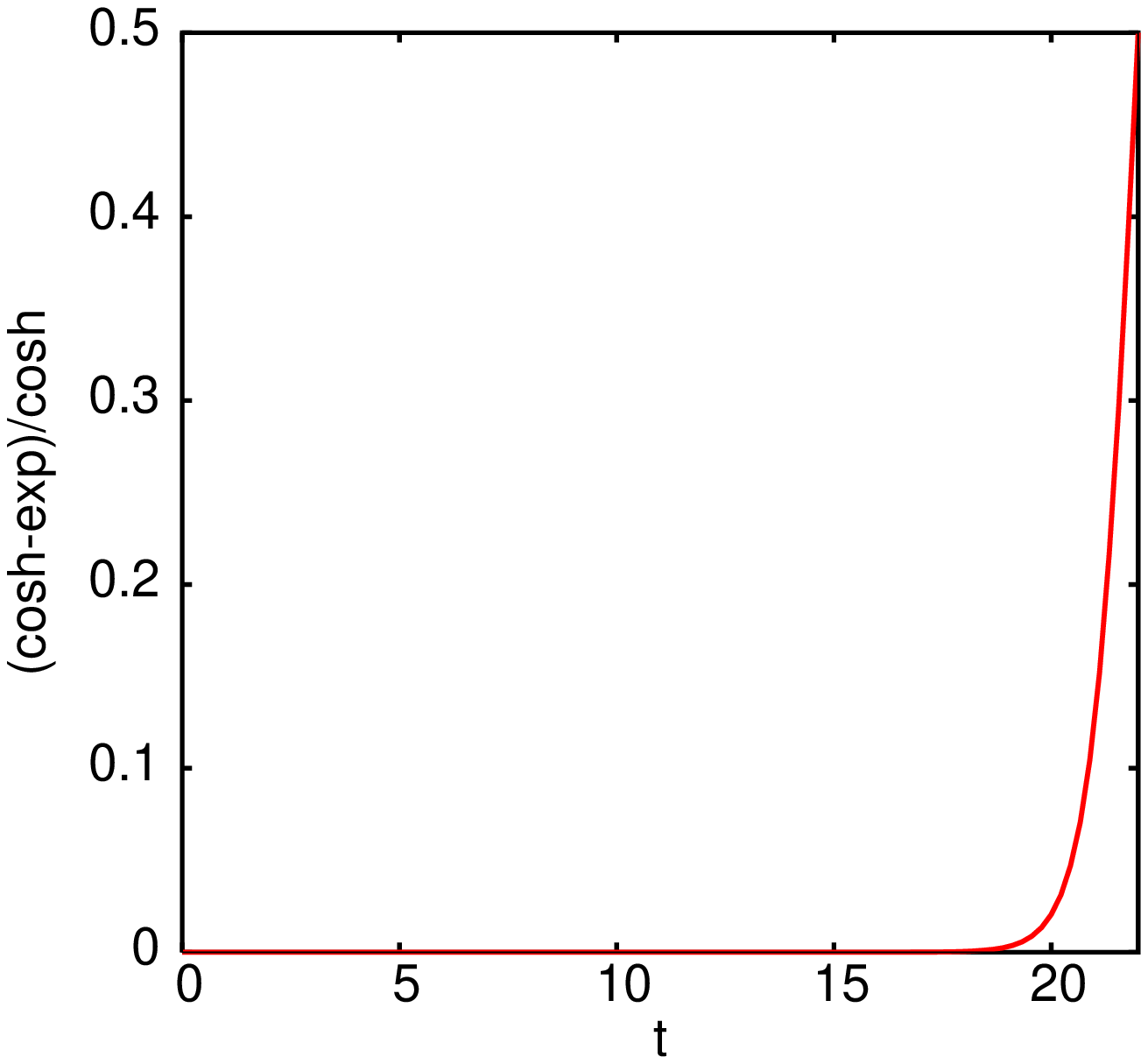}%
  \includegraphics[angle=0,width=0.5\hsize]{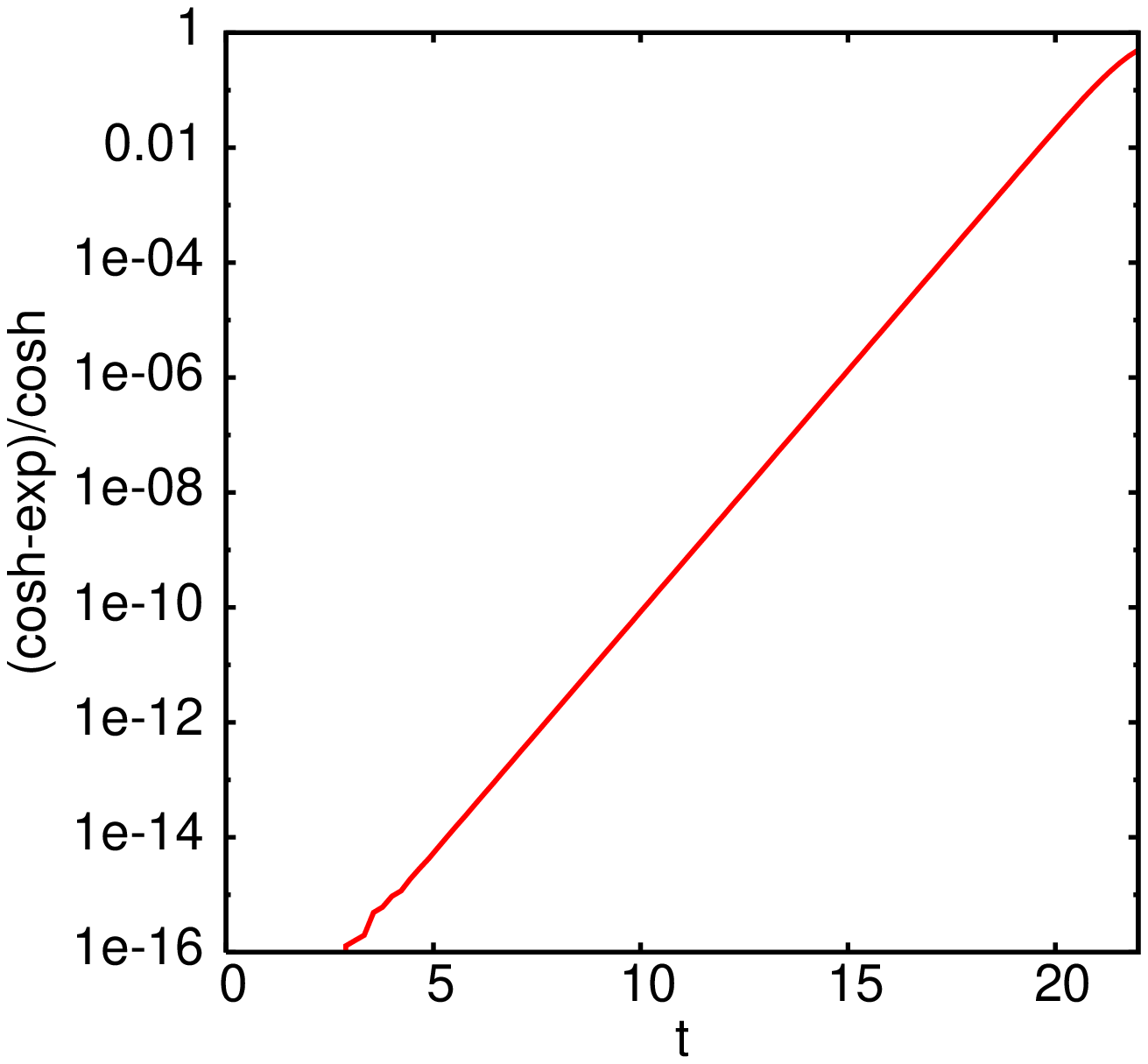}
%
  \caption{\label{Fig:taxis} Data restricted to be along the $t$ axis reveals
           the extent of the wrap-around artifacts (for the $28^3\times 44$
           lattice). A figure of merit $\left(\nu_{\rm cosh}(t)-\nu_{\rm
           exp}(t)\right)/\nu_{\rm cosh}(t)$ is plotted on a linear scale (left
           pane) and on a logarithmic scale (right pane).  A threshold of 1\%
           contamination occurs at two sites from the lattice midpoint.}
  \vspace{0cm}
  \end{center}
\end{figure}

One can then define a figure of merit $\left(\nu_{\rm cosh}(t)-\nu_{\rm
exp}(t)\right)/\nu_{\rm cosh}(t)$.  For the $28^3\times 44$ lattice, this is
plotted (and labeled ``(cosh-exp)/cosh'') in Fig.~\ref{Fig:taxis} with either a
linear (left pane) or logarithmic scale (right pane).  This figure of merit can
be used to identify those times $t$ which are too close to the middle of the
lattice and which suffer substantial contributions from wrap-around effects.

\begin{figure}[ht]
  \vspace{0cm}
  \begin{center}
  \includegraphics[angle=0,width=0.5\hsize]{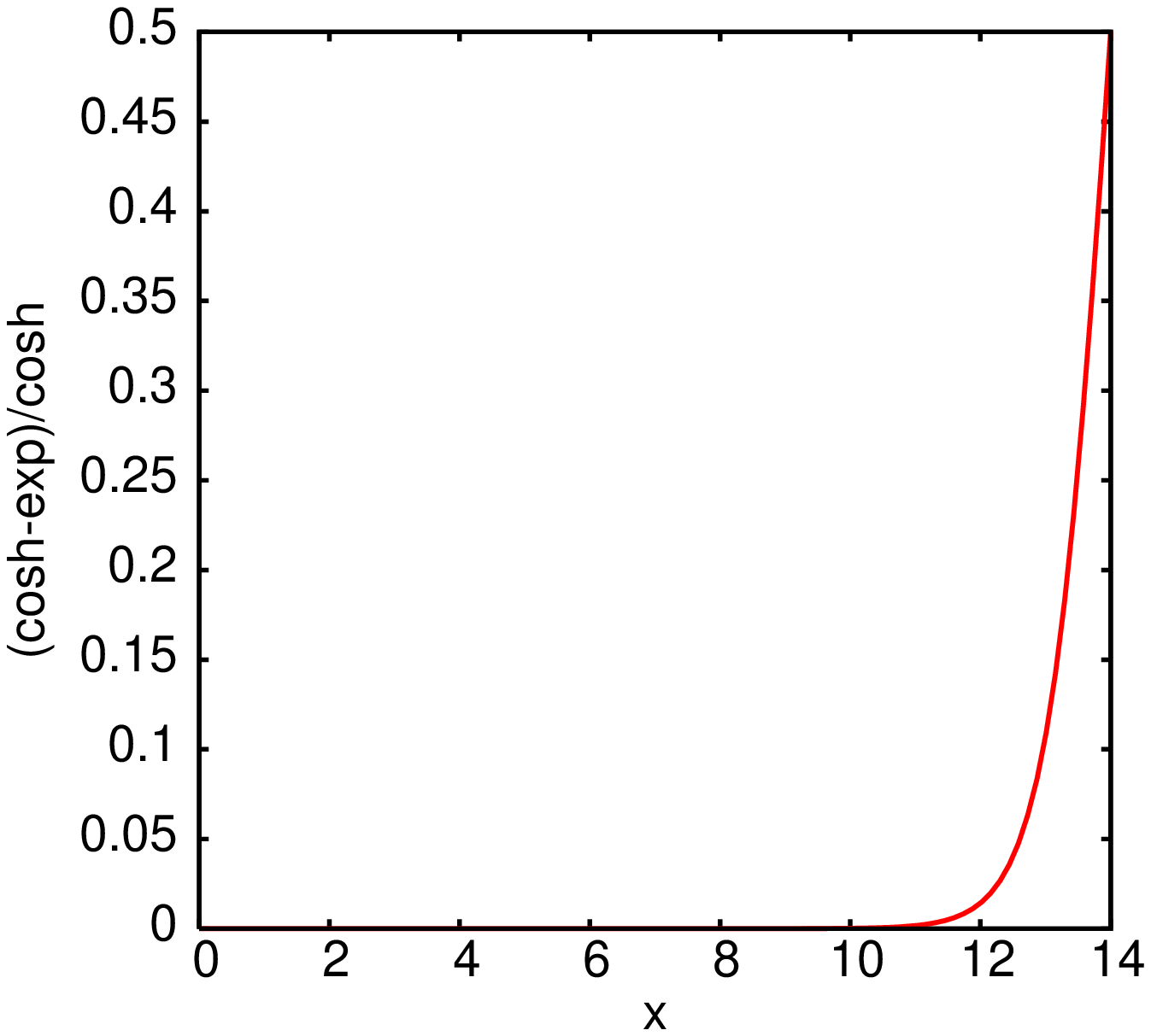}%
  \includegraphics[angle=0,width=0.5\hsize]{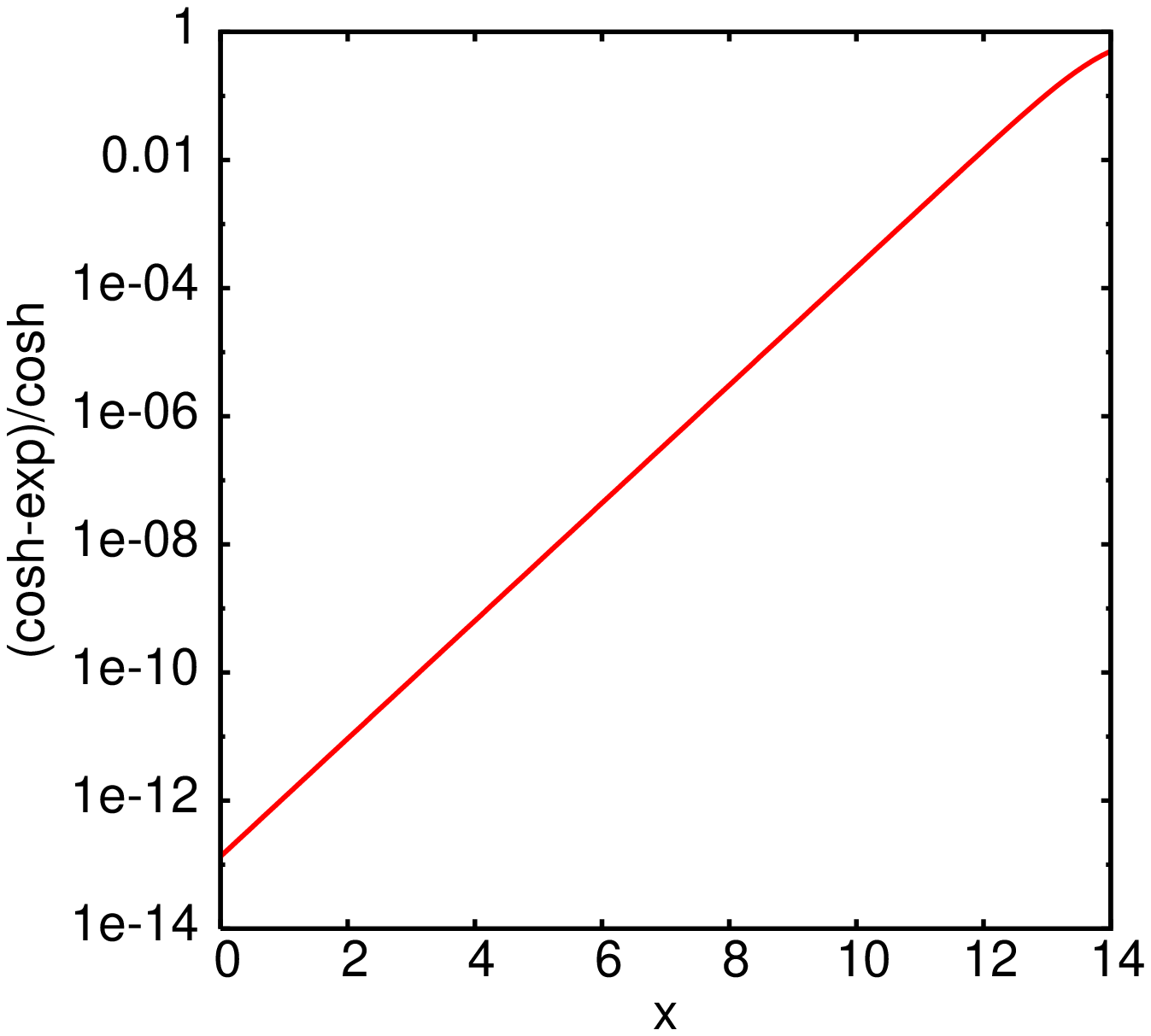}
  \vspace{-0.5cm}
  \caption{\label{Fig:xaxis} Same as for Fig.~\ref{Fig:taxis} but for data
           restricted to be along the $x$ axis.}

  \vspace{0cm}
  \end{center}
\end{figure}

Since our lattices are asymmetric, we repeat the procedure for data restricted
to lie along the $x$ axis; the corresponding figure of merit is plotted in
Fig.~\ref{Fig:xaxis}. These graphs reveal the extent of the wrap-around
artifacts.  To reduce them to a harmless level, we choose to cut the data using
a threshold of 1\% contamination.  From figures Fig.~\ref{Fig:taxis} and
Fig.~\ref{Fig:xaxis}, we see that this occurs at two sites from the lattice
midpoint (for either the $t$ or $x$ direction).

We define $cut=0$ to mean that no data is removed, and likewise define $cut=1$
to mean that data at a point with time coordinate $t$ at the midpoint of the
lattice $t=T/2$ are removed, but data at $|t-T/2|=1,2,\cdots$ are retained.
From Fig.~\ref{Fig:taxis} and Fig.~\ref{Fig:xaxis} then, we find that $cut=3$
(for the $28^3\times 44$ lattice).

Returning the the full data set with lattice points specified by coordinates
$(x,y,z,t)$, we demand that data with {\it any Cartesian component\/} within a
distance of $cut-1=2$ of the midpoint be rejected to ensure that the data that
survives the cut is uncontaminated at about the 1\% level.  Similarly, for the
$20^3\times 32$ lattice, we find that $cut=2$ corresponds to the 1\% threshold.
For the $16^3\times 28$ lattice, $cut=1$.

\begin{figure}[ht]
  \vspace{0cm}
  \begin{center}
  \includegraphics[angle=0,width=0.8\hsize]{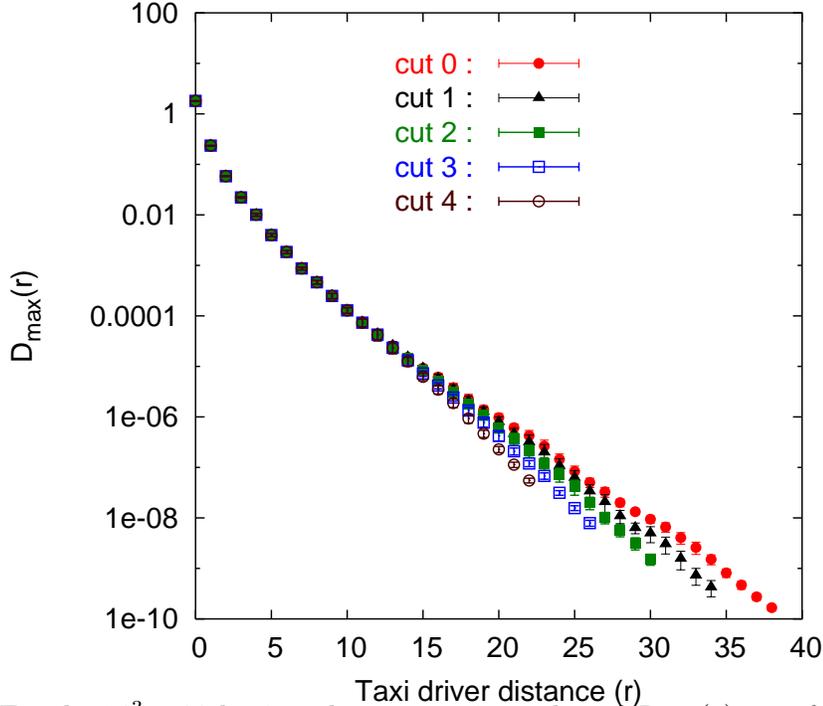}
  \vspace{-0.5cm}
  \caption{\label{Fig:manycuts} For the $16^3\times 28$ lattice, the
           expectation value of $\mathcal{D}_{\rm max}(r)$ as a function of
           taxi-driver distance, $r$, for all data ($cut=0$), and for data
           discarded if within $cut-1$ of any lattice midpoint.  $cut=1$ is
           chosen as our central value wherein wrap-around effects are smaller
           than 1\%.  This is the worst-case scenario; cutting any more would
           reduce the fitted decay distance $r_{\rm ov}$ further.}
  \vspace{0cm}
  \end{center}
\end{figure}

Next, we investigate the consequences of choosing various values for the cut.
As $cut$ increases, more data is discarded and wrap-around effects are further
diminished, but less data is available at large distances from which to extract
the range.  In Fig.~\ref{Fig:manycuts} we plot, for the $16^3\times 28$
lattice, the expectation value of $\mathcal{D}_{\rm max}(r)$ as a function of
taxi-driver distance, $r$, for all data ($cut=0$), and for data discarded if
within $cut-1$ of any lattice midpoint.  We have chosen $cut=1$ as our
preferred value wherein wrap-around effects are smaller than 1\%.  This is a
conservative choice --- the worst-case scenario; cutting any more would reduce
the fitted decay distance $r_{\rm ov}$ further, and runs the risk of increasing
the systematic error of the fitted range due to discarding data in the
large-distance plateau.

\subsection{Locality as Measured by Euclidean Distance}

We conclude that it is perfectly acceptable to simulate overlap fermions with
lattice spacing as coarse as $0.20\,{\rm fm}$ ($f_{\pi}$ scale), since for this
we find that the range is not greater than two lattice units when measured in
taxi-driver distance.  In fact, the situation is even better than it seems.  To
see this, we consider the more familiar standard Euclidean metric
\begin{equation}
  r_{\rm E} \equiv || x - y ||_2 = \sqrt{\sum_{\mu} |x_\mu - y_\mu|^2}
\end{equation}

\begin{figure}[ht]
  \vspace{0cm}
  \begin{center}
  \includegraphics[angle=0,width=0.7\hsize]{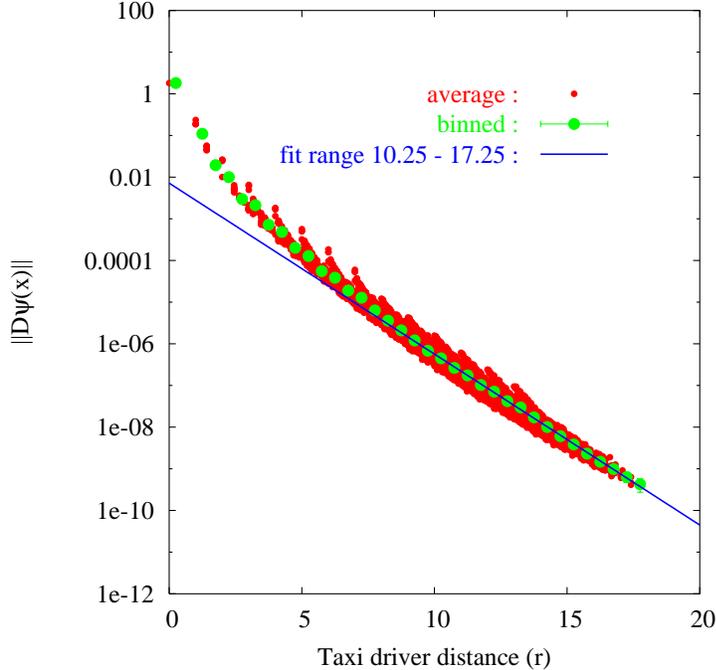}
  \vspace{0cm}
  \caption{\label{Fig:Euclidean} For the $16^3\times 28$ lattice, the
           expectation value of $||D\psi(x)||$ versus Euclidean distance $r$
           (for data cut to remove wrap-around effects).  Data plotted in red
           (darker shade of gray) are averages over configurations; however,
           the maximum has {\it not\/} been taken over points with equal
           Euclidean distance.  The (many fewer) green (lighter shade of gray)
           points are averages of these within bins of width $0.5$.  For large
           $r$, these averages fall exponentially, with range $r_{\rm ov}$.}
  \vspace{0cm} 
  \end{center}

\end{figure}

In Fig.~\ref{Fig:Euclidean} we plot, for our coarsest lattice, $||D\psi(x)||$
(for localized source $\psi_{\alpha}(x)=\delta(x)\delta_{\alpha\beta}$ with
fixed Dirac-color index $\beta$) versus the Euclidean distance.  Here all data
(averaged over configurations) are shown as the profuse red points, {\it
without\/} a maximum taken over points with equal Euclidean distance, in
contrast to Eq.~\ref{Taxi_def}.  Said maxima do not form as smooth an envelope
as in the taxi-driver case; there are many more possible values of Euclidean
distance than taxi-driver distance, which leads to more jitter and sensitivity
to violations of rotational symmetry.  Nevertheless, data are still clearly
contained with a worst-case decay rate.

To obtain an estimate of the range in this case, the data are first smoothed by
averaging over bins of width $0.5$ (in lattice Euclidean distance) to obtain
$\mathcal{D}_{\rm avg}(r)$.  (Taking the maximum over bins, in the spirit of
Eq.~\ref{Taxi_def} results in an envelope which is too sensitive to the choice
of bin size and position.)  The resulting data do have a smooth dependence as a
function of Euclidean distance, and exhibit an exponential tail which is then
fitted.

\begin{figure}[htb]
  \vspace{0cm}
  \begin{center}
  \includegraphics[angle=0,width=0.65\hsize]{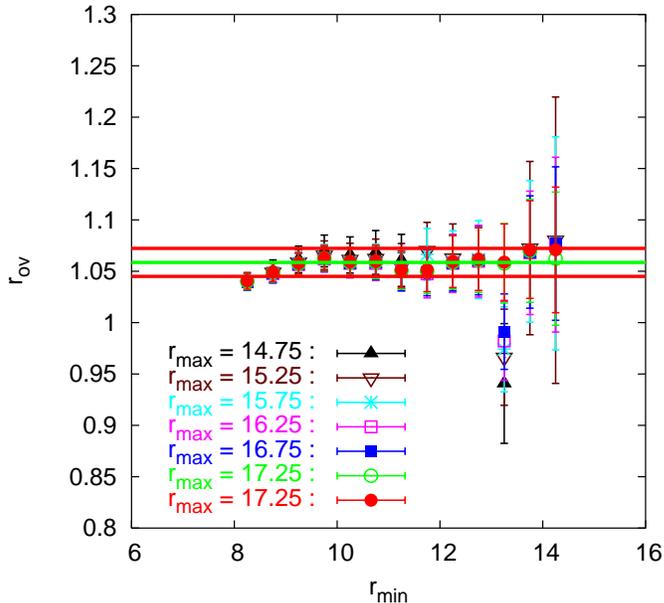}%
  \vspace{0cm}
  \caption{\label{Fig:Euclidean_choice} For the $16^3\times 28$ lattice, the
           fitted values of $r_{\rm ov}$ for various choices of fit interval
           ($r_{\rm min}$,$r_{\rm max}$) are displayed as a function of $r_{\rm
           min}$ for each $r_{\rm max}$, to help choose the best fit.}
  \vspace{0cm}
  \end{center}
\end{figure}

As was done for the taxi-driver case, the fit range is chosen from various
alternatives, as shown in Fig.~\ref{Fig:Euclidean_choice}.  As before, we
choose the least value of $r_{\rm min}$ for which the plateau is clearly and
definitely established, and a value $r_{\rm max}$ for which the fit errors
fairly represent the dispersion of results from various other acceptable fit
intervals.

Choosing a fit range with the above criteria, we fit the tail of the function
with a decaying exponential to extract the range for each of our three
lattices.  We tabulate the results in Table~\ref{Table:Euclidean}.  In
comparing Tables~\ref{Table:locality} and~\ref{Table:Euclidean}, note that
although the two ranges (taxi-driver and Euclidean) differ by a factor of about
two, they are quite compatible heuristically; on a $L^4$ hypercube, the maximum
taxi-driver distance is $4(L/2)=2L$, and the maximum Euclidean distance is
$\sqrt{4(L/2)^2}=L$.

\begin{table}[ht]
  \begin{center}
    \begin{tabular}{ll}
      $a$   &  $r_{\rm ov}$ \\
      \hline
      $0.175\,{\rm fm}$ & $1.06(1)$ \\
      $0.153\,{\rm fm}$ & $0.98(2)$ \\
      $0.113\,{\rm fm}$ & $0.94(6)$ \\
      \hline
    \end{tabular}
    \caption{\label{Table:Euclidean} The range (Euclidean metric) at three
             values of lattice spacing. It is less than about 1 lattice unit
             with lattice spacing as coarse as $0.20\,{\rm fm}$ ($f_{\pi}$
             scale). }
  \end{center}
\end{table}

So even at lattices as coarse as $a=0.20\,{\rm fm}$ ($f_{\pi}$ scale), the
range is about 1 lattice unit (measured using Euclidean distance, or 2 units
using taxi-driver distance).  No unphysical degrees of freedom are induced at
distances longer than the lattice cutoff.

\subsection{A Check: Free Field}

We include here a useful check of our calculations and expectations.  With the
same program used for the interacting-field case, we have calculated $D(x,y)$
for source $y$ and sink $x$ for the free-field case, that is, with all $SU(3)$
matrices on the links set equal to the identity matrix.  We compared the output
with an alternative calculation in terms of Fourier series and found complete
agreement.  As an aside, we then repeated the calculation of the overlap
operator range for the free-field case.

\begin{figure}[ht]
  \vspace{0cm}
  \begin{center}
  \includegraphics[angle=0,width=0.95\hsize]{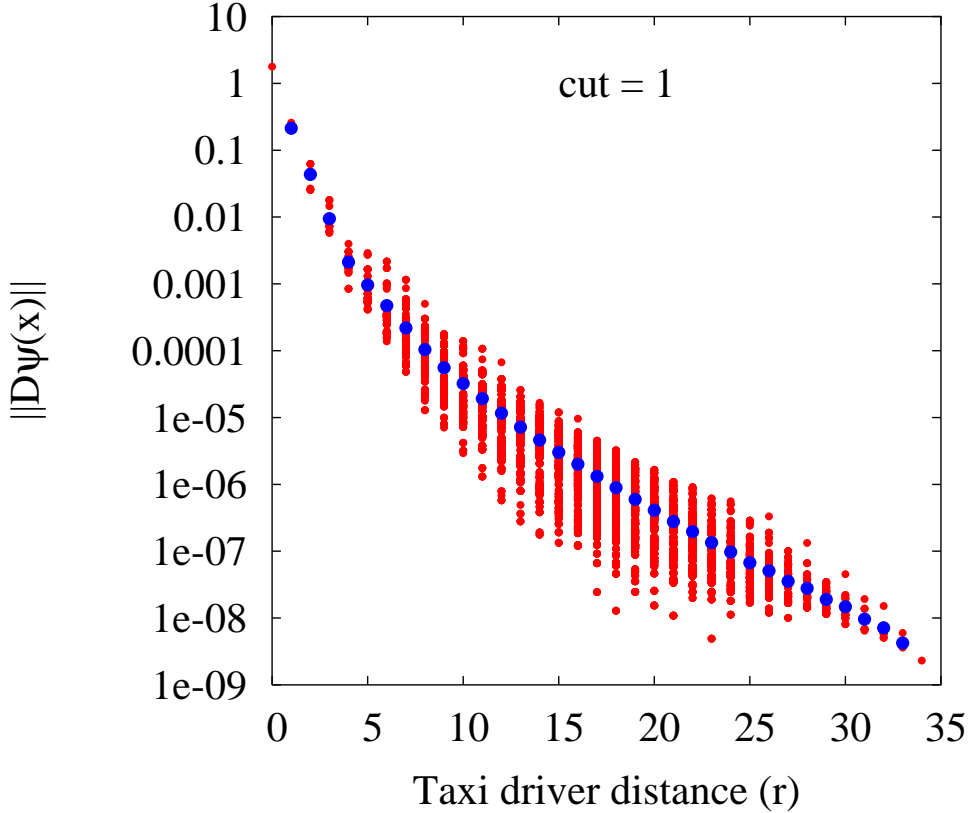}
  \vspace{0cm}
  \caption{\label{Fig:free} Free field case for the $16^3\times 28$ lattice
                 (cut=1). $||D\psi(x)||$ is displayed (red points) for all
                 coordinates; the average $\mathcal{D}_{\rm avg}(r)$ over all
                 equivalent points with a given taxi-driver distance $r$ is
                 shown in blue.}
  \vspace{0.2cm}
  \end{center}
\end{figure}

For the $16^3\times 28$ lattice, we display in Fig.~\ref{Fig:free} a plot for
the free-field case similar to that of the interacting case
(Fig.~\ref{Fig:Taxi_16}), and using the same value to cut the data.  An
important difference, however, is that we display the value of $||D\psi(x)||$
for {\em all\/} points, that is, the maximum is not taken as in
Eq.~\ref{Taxi_def}.  One sees in Fig.~\ref{Fig:free} that, in contrast to the
interacting case, the maximum values for each taxi-driver distance do not form
a smooth envelope.  Presumably, this is a consequence of the lack of rotational
invariance for the free case.  This lack of smoothness makes it difficult to
obtain a quantitative estimate of the range.  A much smoother curve is obtained
by taking the average (mean) $\mathcal{D}_{\rm avg}(r)$, rather than the
maximum, of all equivalent points which have the same taxi-driver distance.
But alas, this is different from what has been presented for the interacting
case.

\begin{figure}[ht]
  \vspace{0cm}
  \begin{center}
  \includegraphics[angle=0,width=0.85\hsize]{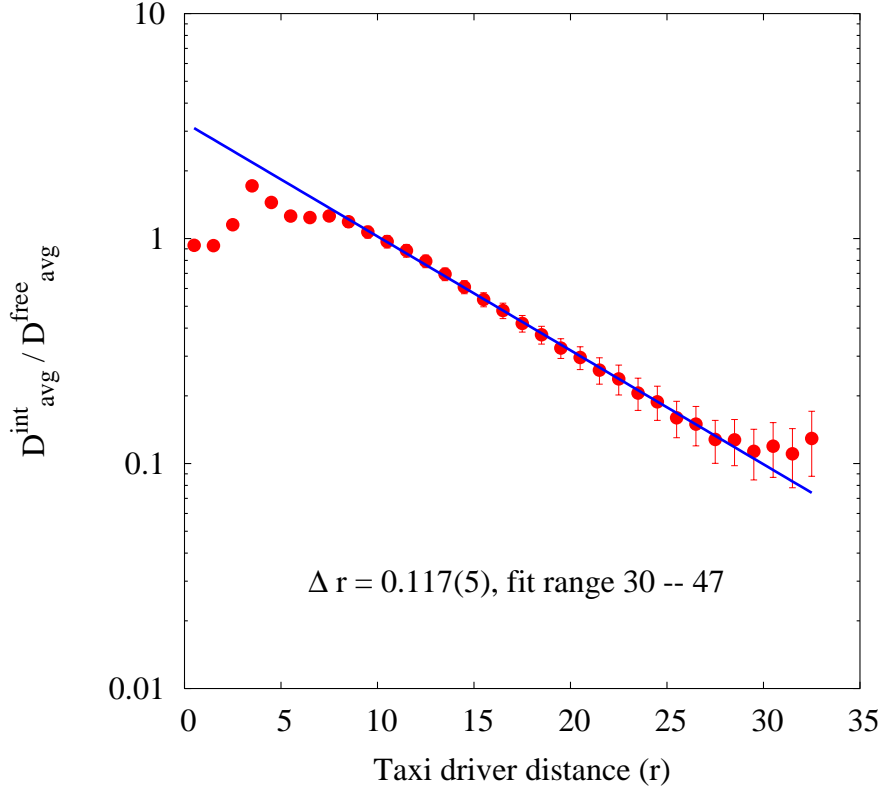}
  \vspace{0cm}
  \caption{\label{Fig:ratio_free} Ratio of interacting to free field case (cut=1).}
  \vspace{0cm}
  \end{center}
\end{figure}

To make a direct comparison of the interacting to the free case, we plot in
Fig.~\ref{Fig:ratio_free} the ratio of $\mathcal{D}_{\rm avg}^{\rm int}(r)$ for
the interacting to $\mathcal{D}_{\rm avg}^{\rm free}(r)$ for the free-field
case.  Here, for each configuration, the average is taken over all coordinates
which have a given taxi-driver distance $r$.  Then an average and standard
error are calculated over all configurations.

For large taxi-driver distance $r$, we expect an exponential decay
$\mathcal{D}_{\rm avg}(r) \sim\exp(-r/r_{\rm ov})$, and thus expect the ratio
to decay exponentially as
\begin{equation}
\mathcal{D}_{\rm avg}^{\rm int}(r) /
\mathcal{D}_{\rm avg}^{\rm free}(r) 
\sim \exp(-r/\Delta r)
\end{equation}
where
\begin{equation}
\frac{1}{\Delta r} \equiv \left( \frac{1}{r_{\rm int}} - \frac{1}{r_{\rm free}}  \right)
\end{equation}
The graph is fit quite nicely with a long straight line giving a value $\Delta
r = 0.117(5)$.  Notice that a positive value indicates that $r_{\rm int} <
r_{\rm free}$.  Initially, this was not expected from preliminary mobility-edge
arguments~\cite{Golterman:2005fe}.  But it is now acknowledged that the
connection between mobility edge and overlap range is severed already in the
free-field case~\cite{Sha06}.  Furthermore, a crude heuristic argument in
support of the observation can be made.  To wit, regard $D(x,y)$ as a sum of
contributions from various paths of links connecting source $y$ to sink $x$.
In the free case, the product of these links is the identity.  In the
interacting case, it is a path-dependent matrix with norm less than one.  Then
one would expect ${\mathcal D}^{int}(r) < {\mathcal D}^{free}(r)$ for each $r$.
If each decays exponentially at large $r$, then one cannot have $r_{\rm int} >
r_{\rm free}$ without contradiction.

\section{Scaling}

At Lattice 2004, Davies {\it et al.}~\cite{Davies:2004hc} collected world data
to demonstrate that different quenched quark formulations could have a
consistent continuum limit.

\begin{figure}[ht]
  \vspace{0cm}
  \begin{center}
  \includegraphics[angle=0,width=0.8\hsize]{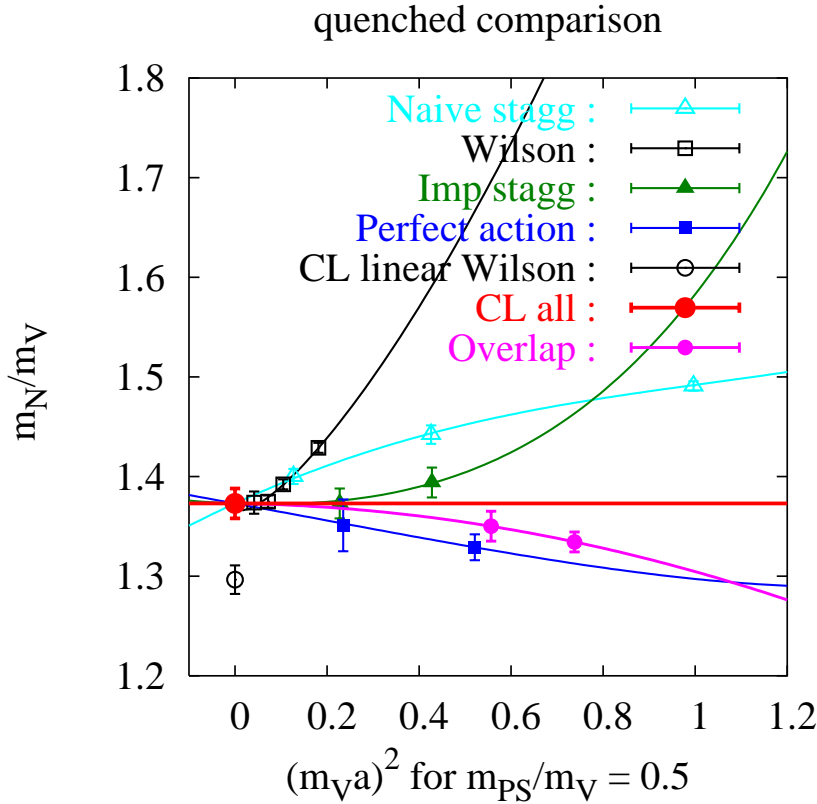}
  \vspace{0cm}
  \caption{\label{Fig:Aoki} ``Aoki'' plot for various quenched data, as
           obtained from~\cite{Davies:2004hc}.  Our data (solid circles) are
           labeled ``overlap''.}
  \vspace{0cm}
  \end{center}
\end{figure}

The conclusion, as illustrated in Fig.~\ref{Fig:Aoki}, is that they could.  But
we emphasize that the constrained fit demanded that there exist a global
continuum limit (by design).  Furthermore, the global continuum limit differed
substantially from the continuum limit obtained solely from the Wilson
formulation where the discretization errors are the largest.  The lesson
learned is that with large discretization errors, it is quite possible to be
misled when extrapolating to the continuum limit, even with high statistics and
many lattice spacings.  It is important to seek a formulation with very small
discretization errors to be able to trust the continuum extrapolation.

Here we add our data ``overlap'' (solid circles in Fig.~\ref{Fig:Aoki}) to the
global quenched spectrum data and find that of all the formulations, its
discretization errors are smallest, allowing for viable computation at
surprisingly coarse lattice spacing.

As another example of the efficacy of the overlap formulation, we made a
non-perturbative computation~\cite{Zhang:2005sc} of the renormalization
constants of composite operators on the $16^{3}\times 28$ lattice using the
regularization independent scheme.

We found that the relations $Z_A=Z_V$ and $Z_S=Z_P$ agree well (within 1\%)
above the renormalization point $p=1.6\,{\rm GeV}$.  The $m\Lambda_{\rm
QCD}a^2$ and $(ma)^2$ corrections of the renormalization are small; the mass
dependence is less than about 3\% up to $ma=0.6$.

\vspace{-0.15cm}

\section{Conclusions}
\label{Conclusions}

\vspace{-0.15cm}

It is viable to simulate quenched overlap fermions at surprisingly coarse
lattice spacing.  Locality is well under control; the range (characteristic
exponential decay length) is about one lattice unit (of Euclidean distance, or
about two lattice units of Taxi-driver distance) for lattice spacing as coarse
as $0.20\,{\rm fm}$ ($f_{\pi}$ scale), such as that used in~\cite{Chen:2003im},
and trends to zero (in physical units) in the continuum limit.  Scaling is
excellent; the Aoki plot is almost flat up to $0.20\,{\rm fm}$.  The overlap
fermion outperforms the other formulations, that is, discretization errors are
the smallest for overlap.  Non-perturbative renormalization of operators show
little mass dependence~\cite{Zhang:2005sc}; e.g.\ less than about 3\% up to
$ma=0.6$ for the renormalization factors.

\section{Acknowledgments}

This work is supported in part by the U.S. Department of Energy under grant
DE-FG05-84ER40154.  The work of NM is supported by U.S. DOE Contract
No. DE-AC05-06OR23177, under which Jefferson Science Associates, LLC operates
Jefferson Lab.








\end{document}